\documentclass[aps,twocolumn]{revtex4}

\newcommand{\ave}[1]{\left\langle #1 \right\rangle}

\usepackage{graphicx}
\begin{document}
\title{Balance Functions, Correlations, Charge Fluctuations and Interferometry}

\author{Sangyong Jeon}  \affiliation{Department of Physics, McGill University,
3600 University St., Montr\'eal, QC H3A-2T8 CANADA\\
RIKEN-BNL Research Center, Brookhaven National Laboratory, Upton, NY 11973}
\email{jeon@hep.physics.mcgill.ca}

\author{Scott Pratt} \affiliation{Department of Physics and
National Superconducting Cyclotron Laboratory\\ Michigan State University,
East Lansing Michigan, 48824} \email{pratt@nscl.msu.edu} 

\date{\today}

\begin{abstract}
\bigskip

Connections between charge balance functions, charge fluctuations and
correlations are presented. It is shown that charge fluctuations can be
directly expressed in terms of a balance functions under certain
assumptions. The distortion of charge balance functions due to experimental
acceptance is discussed and the effects of identical boson interference is 
illustrated with a simple model.

\end{abstract}

\maketitle

\section{Introduction}
\label{sec:intro}

Charge balance functions \cite{bassdanpratt} and charge fluctuations
\cite{Jeon:2000wg,Jeon:2001ka,Ko:2000vp,Asakawa:2000wh} have been proposed as a
means for gaining insight into the dynamics of hadronization in relativistic
heavy ion collisions. Both observables are sensitive to the separation, in
momentum space, of balancing charges. Such a pair is composed of a positive and
negative particle whose charge derives from the same point in space-time. As a
quark-gluon plasma scenario entails a large production of new charges late in
the reaction, a tight correlation between the balancing charge/anti-charge
pairs would provide evidence of the creation of a novel state of matter.

Both balance functions and charge fluctuations can be expressed in terms of
one-particle and two-particle observables. In the following section we present
expressions for both balance functions and charge fluctuations in terms of
spectra and correlations, and show how the charge fluctuations can be simply
expressed in terms of balance functions for neutral systems.

Unfortunately, balance functions and charge fluctuations can both be rather
sensitive to detector acceptance. In Sec. \ref{sec:acceptance} we present a
variant of balance functions which reduces acceptance effects for detectors
with sharp cutoffs in rapidity. Identical pion correlations also affects both
observables in a non-trivial manner. In a simple model utilizing parameters
consistent with observed spectra and correlations from RHIC, we illustrate the
distortion of the balance functions due to Bose-Einstein correlations in
Sec. \ref{sec:hbt}. The insights gained from these studies are summarized in
Sec. \ref{sec:summary}.

\section{Relating Balance Functions, Fluctuations and Correlations}
\label{sec:relatingobservables}

As mentioned in the introduction, both balance-function and charge-fluctuation
observables are generated from one-body and two-body observables which
necessitates that they may be expressed in terms of spectra and two-particle
correlation functions.  In order to express the balance functions in terms of
the elementary correlation functions, first define
\begin{eqnarray}
& & \langle N(a,\Delta_1)\rangle =\int_{\Delta_1} d^3p \frac{dn_a}{d^3p}
\label{eq:NaDef}
\\
{\rm and}\\
\lefteqn{\langle N(b,\Delta_2;a,\Delta_1)\rangle }\\
\nonumber
& = &
\int_{\Delta_1} d^3p_a\int_{\Delta_2}d^3p_b
\frac{d^2n_{ab}}{d^3p_ad^3p_b}
\label{eq:NabDef}\\
&=&
\int_{\Delta_1} d^3p_a\int_{\Delta_2}d^3p_b
\frac{dn_a}{d^3p_a}\frac{dn_b}{d^3p_b} C(b,\Delta_2; a,\Delta_1)
\end{eqnarray}
where $\Delta_{1,2}$ are phase space criteria such as rapidity intervals.

In terms of these quantities, the balance function is expressed as
\begin{eqnarray}
\label{eq:balancedef}
B(\Delta_2|\Delta_1)&=&\frac{1}{2}\left\{
D(-,\Delta_2|+,\Delta_1) 
- 
D(+,\Delta_2|+,\Delta_1)\right.
\\
\nonumber
&&
+
\left.
D(+,\Delta_2|-,\Delta_1) 
- 
D(-,\Delta_2|-,\Delta_1)
\right\}.
\end{eqnarray}
where
\begin{equation}
D(b,\Delta_2|a,\Delta_1)
=\frac{\langle N(b,\Delta_2;a,\Delta_1)\rangle}{\langle N(a,\Delta_1)\rangle},
\end{equation}
which can be considered as a conditional probability.

Thus correlation functions and spectra are sufficient to determine balance
functions, although the required integration can be somewhat convoluted,
depending on the binning, $\Delta_1$ and $\Delta_2$.  The criteria $\Delta_1$
is based solely on the momenta of the first particle, while the criteria
$\Delta_2$ might be any function of the momenta of both particles, e.g., it
might be determined by the relative rapidity of the two particles. 

To establish the correspondence between charge fluctuations and balance
functions, consider a balance function binned as a function of the rapidity
difference where both particles are required to reside within a fixed rapidity
window of size $Y$.  For this case $\Delta_1$ constrains the first particle to
be within the rapidity window, and $\Delta_2$ constrains the second particle
to have a relative rapidity $|y_b-y_a|=\Delta y$ while also existing inside the
rapidity window. This binning was applied in preliminary results from STAR
reported in \cite{tonjes_parkcity}. Referring to this balance function as
$B(\Delta y|Y)$, one can find the charge fluctuation within the rapidity window
$0 < y < Y$ by integrating $B(\Delta y|Y)$ in the interval $0<\Delta y<Y$.  In
this case, 
\begin{eqnarray}
B(Y|Y)&=&\lefteqn{\int_0^Y d\Delta y~B(\Delta y | Y) }\\
\nonumber
& = &{1\over 2}\left\{
\frac{\left\langle{N_+N_-}\right\rangle_\Delta}{ 
\left\langle{N_+}\right\rangle_\Delta}
+\frac{\left\langle{N_+N_-}\right\rangle_\Delta}
{\left\langle{N_-}\right\rangle_\Delta}  \right.\\
\nonumber
& & {}\left.
-\frac{\left\langle{N_+(N_+-1)}\right\rangle_\Delta}
{\left\langle N_+\right\rangle_\Delta}
-\frac{\left\langle N_-(N_--1)\right\rangle_\Delta}
{\left\langle{N_-}\right\rangle_\Delta}
\right\}
\end{eqnarray}
where $\langle\cdots\rangle_\Delta$ denotes averages in the phase space region
$\Delta$.
Writing $N_{\pm} = \langle N_\pm\rangle_\Delta + \delta N_\pm$,
it is not hard to show
\begin{eqnarray}
\frac{\langle(Q-\langle Q\rangle)^2\rangle}{\langle N_{\rm ch}\rangle}
=1-\int_0^Y d\Delta y~B(\Delta y | Y)
+O\left(\frac{\langle Q\rangle}{\langle N_{\rm ch}\rangle}\right)
\label{eq:relation}
\end{eqnarray}
where $Q = N_+ - N_-$ and $N_{\rm ch} = N_+ + N_-$. For electric charge, the
size of the correction is usually less than 5\,\% in relativistic heavy ion
collisions where the number of produced charges is much greater than the net
charge. However, for baryon number the additional term is not negligible even
at RHIC.

In a boost-invariant system (independent of rapidity) the balance function
$B(\Delta y|Y)$ can be related to the balance function for an infinite
interval.
\begin{equation}
\label{eq:binsizecorrection}
B(\Delta y|Y)=B(\Delta y|Y=\infty)(1-\Delta y/Y).
\end{equation}
The factor $(1-\Delta y/Y)$ accounts for the probability that a particle's
partner will fall within the rapidity window given that they are separated by
$\Delta y$. Also, assuming boost invariance allows one to express the balance
functions simply in terms of correlation functions as described in
Eq. (\ref{eq:balancedef}).
\begin{eqnarray}
B(\Delta y|Y=\infty)
&=&\frac{1}{2}\left\{\frac{dn_+}{dy}C_{++}(\Delta y)\right.
\\
\nonumber
&&\hspace*{-50pt}\left.+\frac{dn_-}{dy}C_{--}(\Delta y)
-\left(\frac{dn_+}{dy}+\frac{dn_-}{dy}\right)C_{+-}(\Delta y)\right\}.
\end{eqnarray}

From the above discussion it is clear that the charge fluctuation is the global
measure of the charge correlation and the balance function is a differential
measure of the charge correlation and therefore carries more information. The
advantage of charge fluctuations is that they carry a clear physical meaning in
terms of a grand canonical ensemble ~\cite{Jeon:2001ka,Asakawa:2000wh}, and can
therefore be easily connected to more ideal theoretical models, e.g. Lattice
QCD calculations.  However, since there are no external sources of charge in
heavy ion collisions to warrant a grand canonical treatment, both observables
are effectively driven by the dynamics of how balancing charges are formed and
separate.

We emphasize here that charge fluctuations were not intended to provide a
derivative measure.  As can be seen from Eq. (\ref{eq:relation}) the charge
fluctuation summarizes the balance functions in one number. It gives somewhat
different information than the width of the balance function since it is also
affected by the height. We do not recommend analyzing charge fluctuations as a
function of the size of the rapidity window. If the different sized windows
included the same pairs, the values would no longer be statistically
independent when plotted against the window size. If the windows are used only
once, the information from pairs which occupy adjacent windows is thrown away.

A similar set of issues surfaced in making the connection between fluctuations
and correlations in the study of multiplicity distributions analyzed as a
function of rapidity \cite{wieandprattbalantekin,haglinseibert}. A more general
connection between fluctuation and inclusive observables can be found in
\cite{Bialas:1999tv}. However, it should be noted that factorial moments and
scaled factorial moments, which are measures of fluctuation
\cite{bialaspeschanski,seibertsfm}, offer the opportunity to study $n$-body
correlations for $n>2$ in a manner which, unlike correlations, can be easily
collapsed into a single variable.

\section{Minimizing Acceptance Effects in Balance Functions}
\label{sec:acceptance}

Balance functions analyzed by the STAR collaboration \cite{tonjes_parkcity}
were constructed according to the prescription that $p_1$ would refer to any
pion that is measured within a specified rapidity window while $p_2$ referred
to the relative rapidity, again with the requirement that the second particle
was within the rapidity window.  In that case,
\begin{eqnarray}
\label{eq:oldnormexample}
B(\Delta y|Y)&=&\frac{1}{2}\left\{
\frac{\ave{N_{+-}(\Delta y)}-\ave{N_{++}(\Delta y)}}{\ave{N_+}}\right.\\
\nonumber
&&\hspace*{32pt}+\left.\frac{\ave{N_{-+}(\Delta y)}-\ave{N_{--}(\Delta y)}}
{\ave{N_-}}
\right\}.
\end{eqnarray}
Here $N_{+-}(\Delta y)$ counts pairs with opposite charge that satisfy the
criteria that their relative rapidity equals $\Delta y$, whereas $N_+$ is the
number of positive particles in the same interval.  Here the angular bracket
represents averaging over the events and $Y$ is the size of the detector
rapidity window. From this example, one can readily understand how balance
functions identify balancing charges. For any positive charge, there exists
only one negative particle whose negative charge derived from the point at
which the positive charge was created. By subtracting from the numerator the
same object created with positive-positive pairs, one is effectively
subtracting the uncorrelated negatives from the distribution and identifying
the balancing charge on a statistical basis.

From the construction of Eq. (\ref{eq:oldnormexample}), one can understand the
sensitivity of $B(\Delta y|Y)$ to the acceptance size $Y$ by considering a
detector which covers a finite range in rapidity.
\begin{equation}
y_{\rm min}<y<y_{\rm max}.
\end{equation}
with $y_{\rm max} - y_{\rm min} = Y$.  For this example, the balance function
must go to zero as $\Delta y$ approaches $Y$. This occurs because the particle
satisfying the condition $p_1$ must lie at the extreme boundary of the
acceptance in order for the second particle to have a relative rapidity $\Delta
y\sim Y$ while remaining in the acceptance. The balance function is thus forced
to zero at the limits of the acceptance for trivial reasons.

Of course, the balance function corresponding to a perfect detector, $B(\Delta
y|\infty)$, is independent of any particular detector size $Y$. As described in
Eq. (\ref{eq:binsizecorrection}), one can easily correct for the detector
acceptance in the boost-invariant case by dividing the balance function by
a factor $(1-\Delta y/Y)$.

These balance functions would not have more information than those created
without the correction factor, but the information would more directly address
the physics of charge separation rather than reflecting the experimental
acceptance. We note that the statistical uncertainties of the corrected balance
function will however be quite large as $\Delta y\sim Y$.

More generally, one can correct the balance functions for the acceptance by
dividing the numerators in Eq. (\ref{eq:balancedef}), $N(Q_2,p_2|Q_1,p_1)$, by
acceptance factors, $A(Q_1,p_2|Q_1,p_1)$.
\begin{eqnarray}
B(p_2|p_1)&=&\frac{1}{2}\left\{
\frac{N(-,p_2|+,p_1)}{A(-,p_2|+,p_1)N(+p_1)}\right.\\
\nonumber
&&\hspace*{-55pt}-\frac{N(+,p_2|+,p_1)}{A(+,p_2|+,p_1)N(+,p_1)}
+\frac{N(+,p_2|-,p_1)}{A(+,p_2|-,p_1)N(-,p_1)}\\
\nonumber
&&\left.\hspace*{55pt}-\frac{N(-,p_2|-,p_1)}{A(-,p_2|-,p_1)N(-,p_1)}
\right\}.
\end{eqnarray}
The acceptance factor represents the probability that, given a particle $i$
satisfied the criteria $p_1$, a second particle that satisfied $p_2$ would be
detected.  Since the criteria $(Q_2,p_2)$ may depend on the individual particle
that satisfied $(Q_1,p_1)$, it may be simpler to calculate $A$ in terms of
$a_i(Q_2,p_2)$ which represents the acceptance into $(Q_2,p_2)$ given the
particular particle $i$.
\begin{equation}
A(Q_2,p_2|Q_1,p_1)=\frac{\sum_{i\in Q_1,p_1} a_i(Q_2,p_2)}{N(Q_1,p_1)}.
\end{equation}
The acceptance probability $a_i(Q_2,p_2)$ would be between zero and unity. We
note that the acceptance is effectively accounted for by performing a
substitution for the denominators in Eq. (\ref{eq:balancedef}).
\begin{equation}
N(Q_1,P_1)\rightarrow \sum_{i\in Q_1,p_1} a_i(Q_2,p_2)
\end{equation}

For the boost-invariant case above where $p_2$ referred to the relative
rapidity, and where the acceptance is represented by simple step functions in
rapidity, the probabilities would become
\begin{equation}
a_i(p_2)=\left\{
\begin{array}{cl}
0,& y_{\rm max}-y_i<\Delta y{~\rm and~} y_i-y_{\rm min}<\Delta y\\
1/2,&y_{\rm max}-y_i<\Delta y{~\rm and~} y_i-y_{\rm min}>\Delta y\\
1/2,&y_{\rm max}-y_i>\Delta y{~\rm and~} y_i-y_{\rm min}<\Delta y\\
1,&y_{\rm max}-y_i>\Delta y{~\rm and~} y_i-y_{\rm min}>\Delta y
\end{array}
\right.
\end{equation}
Given that the bins $p_2$ would be of finite extent, the values might differ
from $1/2$, $0$ or unity if the bin straddled the acceptance. For a boost
invariant system, averaging over $y_i$ results in the simple correction factor
$(1-\Delta y/Y)$ mentioned previously.

In general, if the acceptance depends on where in the $(Q_2,p_2)$ bin the
second particle lies, one can not calculate the acceptance correction exactly
without knowledge of the charge correlation which is unavailable except by
measuring the balance function in sufficiently small $p_2$ bins such that the
acceptance is effectively uniform within the small bins. This may not be
feasible due to statistics. It is our recommendation that such factors
$a_i(Q_2,p_2)$ should be kept simple. One can always correct theoretical
results for the detector response by applying whatever factor is applied to the
experimental analysis. Although comparisons with models could have been made
without any corrections, acceptance-corrected balance functions can allow for a
more physical interpretation while not compromising the integrity of the
analysis.

\section{Bose-Einstein Correlations and Balance Functions}
\label{sec:hbt}

Although Bose-Einstein correlations only affect identical particles at small
relative momentum, they manifest themselves in balance functions despite the
fact that the binning in balance functions typically covers a large volume in
momentum space. In a related topic, Bose-Einstein correlations (also known as
the Hanbury-Brown Twiss effect, HBT) have been observed in rapidity
correlations where all charged particles, both positive and negative, were used
in the analysis \cite{emu01,tannenbaum,wieandprattbalantekin}. The
manifestations of HBT in balance function derives from the fact that it induces
a correlation between a given charge and all other charges, not just those that
were created to balance the given charge.

In balance functions the HBT effect should enhance the probability that
same-charge particles have small relative momentum, thus providing a dip in the
balance function at small relative rapidity. In order to model this effect, we
consider pairs of pions with momenta $p_a$ and $p_b$ and opposite charge that
are created according to a boost-invariant thermal distribution with a
temperature of 190 MeV, thus roughly reproducing the pion spectra measured in
Au + Au collisions at RHIC. In addition to the usual contribution to the
balance function between $p_a$ and $p_b$, a second component derives from the
interaction with other pions from other pairs which in this case have momenta
$p_c$ and $p_d$. The thermal distribution describing the first two particles
was centered at zero rapidity, while the thermal distribution responsible for
emission of the second pair was randomly chosen within $\pm 4$ units.

The particles $p_a$ and $p_c$ were assumed to have the same sign, as were the
particles with momenta $p_b$ and $p_d$. A contribution to the balance function
was constructed using these particles, but with a weight,
\begin{equation}
w=C_{++}(p_a,p_c)C_{--}(p_b,p_d)C_{+-}(p_a,p_d)C_{-+}(p_b,p_c).
\end{equation}
This accounts for the weight due to two-particle interactions. The correlation
functions were simple functions of $Q_{\rm inv}(p_a,p_b)\equiv
\sqrt{(p_a-p_b)^2}$, which were generated by calculating correlation functions
for a spherically symmetric Gaussian source of radius, $R_{\rm inv}= 7$ fm,
again crudely in line with measurements at RHIC\cite{starhbt}. The weights were
calculated by averaging the squared relative wave function for two particles,
including the Coulomb interaction between the pions. The weight was multiplied
by the number of such pairs which came from assuming that there were 200 pion
pairs per unit rapidity. Only a fraction, $\lambda=0.7$, of the pairs were
assumed to interact due to the fact that some pions would be created in
long-lived decays. The acceptance of the STAR detector, and the fact that only
a fraction of the pions would truly be balanced by other pions (rather than by
charged kaons or other particles) was roughly accounted for by accepting only
60\% of the particles with transverse momenta between 100 MeV/c and 700 MeV/c.

\begin{figure}
\centerline{\includegraphics[width=0.45\textwidth]{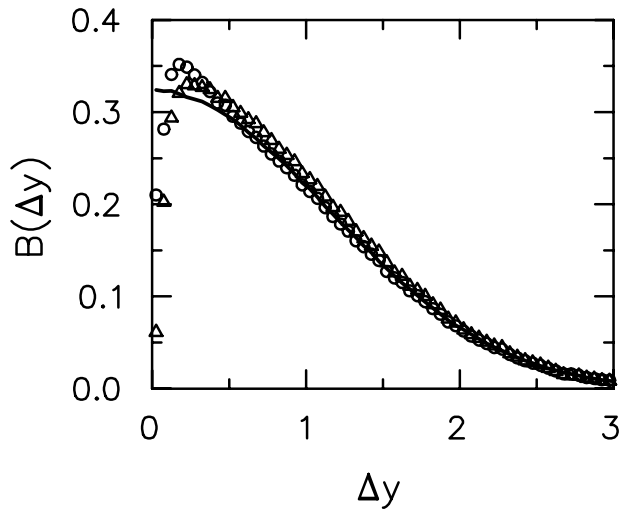}}
\caption{
\label{fig:hbt}
The balance function from the simple thermal Bjorken model (line) has been
parameterized and filtered to roughly provide rough consistency with
measurements of STAR. The inclusion of HBT effects (triangles) gives a dip at
small $\Delta y$, while the extra addition of Coulomb (circles) modifies the
dip.}
\end{figure}

The resulting balance functions are displayed in Fig. \ref{fig:hbt}. When the
interactions between particles is neglected, the resulting balance function
falls monotonically, and has a width consistent with the temperature. The
inclusion of the HBT effect results in a large dip near $\Delta y=0$, and an
enhancement at somewhat larger $\Delta y$. The dip derives from the enhancement
of same-sign pairs which results in a negative contribution to the balance
function. Since the weight is assigned to the emission of the $cd$ pair, the
positive HBT weight contributes to opposite sign pairs with equal strength, but
is spread out over a wider range of $\Delta y$. Hence, the balance function is
slightly enhanced for $\Delta y\sim 1/2$ from HBT effects.

Also shown in Fig. \ref{fig:hbt} are calculations where the Coulomb interaction
is included. Since Coulomb interactions result in attractions for opposite-sign
pairs, and repulsions for same sign pairs, the dip due to HBT is mitigated.

Although the shape of the balance function is visibly altered by the inclusion
of two-particle interactions, the mean width changed by only a few percent. The
strength of the distortion was proportional to the multiplicity, but the effect
is not necessarily weaker for peripheral events. This follows because the HBT
correction contributes with a strength proportional to $R^{-3}$. Since the
product of $dn/dy$ and $R^{-3}$ stays roughly constant over a wide range of
centralities in heavy ion collisions, the distortion due to interactions should
not appreciably affect the centrality dependence of the balance function's
width.

\section{Summary}
\label{sec:summary}
This paper covered several technical issues related to balance functions.  The
conclusion of Sec. \ref{sec:relatingobservables} is that charge fluctuations
can be related to balance functions in a straightforward manner unless the
average net charge is large. In fact, the charge fluctuation can be thought of
as a measure of the integrated balance function $B(\Delta y|Y)$ from zero to
$Y$. In that sense, it represents a one-component measure of the balance
function, just like the mean width.

Section \ref{sec:acceptance} provided an illustration of how balance functions
can be created in such a way as to minimize sensitivity to experimental
acceptance. Although the example addressed only problems with finite acceptance
in rapidity where the balance function was binned according to relative
rapidity, the principles could be applied to balance functions in any
variables.

The last section considered the inclusion of two-particle interactions into
Balance functions. The effects were shown to be quite visible at small relative
rapidity. However, the width of the balance function was not significantly
affected by the two-particle interactions. This is encouraging, as it justifies
interpreting balance functions as objects that statistically identify balancing
partners, while subtracting out contributions from other pairs.

\acknowledgments{This work was supported by the National Science Foundation,
Grant No. PHY-00-70818, by the Natural Sciences and Engineering Council of
Canada and by le Fonds pour la Formation de Chercheurs et l'Aide \`a la
Recherche du Qu\'ebec.}

\bibliographystyle{prsty}
\bibliography{balance}

\end{document}